\documentclass[12pt]{article}

\usepackage{scicite}
\usepackage{times}
\usepackage{graphicx}
\usepackage{gensymb}
\usepackage{amsmath}
\usepackage{braket}

\newenvironment{sciabstract}{%
\begin{quote} \bf}
{\end{quote}}

\title{Reactions Between Layer-Resolved Molecules Mediated by Dipolar Exchange}

\author
{William G. Tobias$^{1\ast}$, Kyle Matsuda$^{1}$, Jun-Ru Li$^{1}$, Calder Miller$^{1}$,\\
Annette N. Carroll$^{1}$, Thomas Bilitewski$^{1}$, Ana Maria Rey$^{1}$, Jun Ye$^{1\ast}$
\\
\normalsize{$^{1}$
JILA, National Institute of Standards and Technology,}\\
\normalsize{and Department of Physics, University of Colorado,}\\
\normalsize{Boulder, CO 80309}
\\
\normalsize{$^\ast$Corresponding authors; e-mails: william.tobias@colorado.edu, ye@jila.colorado.edu}
}

\date{}

\begin{document} 

\baselineskip24pt

\maketitle 

\begin{sciabstract}

Microscopic control over polar molecules with tunable interactions would enable realization of novel quantum phenomena. Using an applied electric field gradient, we demonstrate layer-resolved state preparation and imaging of ultracold potassium\nobreakdash-rubidium molecules confined to two-dimensional planes in an optical lattice. The coherence time of rotational superpositions in individual layers is maximized by rotating the electric field relative to the optical trap polarization to achieve state-insensitive trapping. Molecules in adjacent layers interact via dipolar exchange of rotational angular momentum; by adjusting the interaction strength between spatially separated ensembles of molecules, we regulate the local chemical reaction rate. The observed resonance width of the exchange process vastly exceeds the dipolar interaction energy, an effect we attribute to the thermal energy. This work realizes precise control of interacting molecules, enabling electric field microscopy on subwavelength length scales and allowing access to unexplored physics in two-dimensional systems.

\end{sciabstract}

\section*{Main Text}

Ultracold polar molecules, which have complex internal structures and dipole moments tunable with external electric fields, represent a model system for studying many-body physics\cite{Micheli2006,Goral2002,Baranov2012,Bohn2017}. In reduced dimensionality, the sign and magnitude of interactions between molecules depend on the orientation of the dipole moments with respect to the external confinement. Within two-dimensional (2D) layers, for example, the averaged interactions between molecules can be varied continuously from attractive to repulsive by rotating the dipoles into and out of the plane. Molecules in an isolated 2D layer are predicted to exhibit diverse quantum phenomena determined by the dipole angle and other parameters including electric field and rotational state. These include complex ground state phases such as superfluids and topological insulators\cite{Yao2013,Gorshkov2011_2,Manmana2013,Yao2018,Peter2012,Syzranov2014,Cooper2009,Levinsen2013}, collective excitations in the hydrodynamic regime\cite{Babadi2012}, and interaction-enhanced rotational coherence and dynamical generation of spin squeezing\cite{Bilitewski2021}. Molecules prepared in multiple 2D layers may pair and form states with long-range order\cite{Wang2006,Potter2010,Pikovski2010}. Addressing individual lattice layers would allow initialization of varied configurations to realize these models: single layers, where molecules are isolated against out-of-plane interactions, and minimal systems with interlayer interactions such as bilayers and trilayers (two and three adjacent layers, respectively).

Recent experimental progress with molecules in 2D has included reaching quantum degeneracy using direct evaporation\cite{Valtolina2020} or pairing in a degenerate atomic gas\cite{Zhang2021}, performing optical microscopy of single lattice sites\cite{Rosenberg2021}, and lengthening the coherence time of rotational superpositions\cite{Neyenhuis2012,Seebelberg2018}. When translational motion is allowed within layers, as is the case for confinement in a 1D optical lattice, molecules approaching at short range undergo lossy chemical reactions\cite{Ospelkaus2010,Hu2019,Bause2021,Gersema2021,Gregory2020}. These losses can be mitigated by orienting the dipole moments perpendicular to the layer\cite{Quemener2010,Ni2010,DeMiranda2011,Valtolina2020,Frisch2015} or engineering rotational state couplings\cite{Matsuda2020,Li2021,Anderegg2021,Wang2015,Karman2018} to generate a repulsive collisional barrier. A major missing capability is the ability to prepare molecules in different internal states and control multiple layers individually, which is essential for tuning dipolar interactions in reduced dimensionality.

Here, we demonstrate layer-resolved imaging and rotational state preparation of ultracold potassium-rubidium (KRb) molecules by applying an electric field gradient to shift rotational transitions between lattice layers. This method is inspired by previous works with atoms\cite{Karski2010,Sherson2010,Edge2015}. Using the capability to address single layers, we characterize the dynamics of highly controlled few-layer systems interacting by exchanging rotational angular momenta (``spin exchange"), a process mediated by dipolar interactions\cite{Yan2013}. Within individual layers, molecules experience both long-range dipolar interactions and short-range chemical reactions, while separated layers only interact via dipolar spin exchange. In the case of multiple layers containing different rotational states, spin exchange leads to mixing of rotational state populations, which strongly enhances the chemical reaction rate\cite{Pikovski2011}. Local control of molecule layer occupation and internal state allows us to probe these dynamics and will enable the exploration of quantum phases in two dimensions.

Starting with degenerate $^{40}$K and $^{87}$Rb atoms in an optical dipole trap (ODT), we load a 1D optical lattice to form a stack of 2D layers, with an interlayer spacing $a=540$ nm\cite{Valtolina2020}. Using magnetoassociation and stimulated Raman adiabatic passage (STIRAP) at a bias electric field $\left|\textbf{E}\right|=1$ kV/cm, we associate the atoms into KRb\cite{Ni2008,DeMarco2019}. KRb is formed in the rotational ground state $\ket{0}$, where $\ket{N}$ denotes the state with electric field-dressed rotational quantum number $N$ and zero angular momentum projection ($m_N=0$) onto the quantization axis specified by \textbf{E}. The harmonic trapping frequencies for $\ket{0}$ in the combined trap are $(\omega_x,\omega_y,\omega_z)=2\pi\times(42,17\,000,48)$ Hz. The molecules are pinned along \textbf{y}, parallel to gravity, but are free to move radially (\textbf{x}-\textbf{z} plane, Fig. 1A). For the typical temperature $T=350$ nK, $k_B T=0.4\,\hbar\omega_y$, where $k_B$ is the Boltzmann constant and $\hbar=h/2\pi$ is the reduced Planck constant, so the molecules occupy only the lowest lattice band. Due to the thermal extent of the atomic clouds along \textbf{y} prior to loading the optical lattice, the initial molecule distribution spans approximately 12 lattice layers, with a peak of about 1500 molecules per layer. Compared with our previous work in 2D\cite{Valtolina2020}, where an auxiliary optical trap was used to compress KRb into few lattice layers to increase peak density, we deliberately prepare a broad distribution of molecules to minimize population differences among the central layers prior to layer addressing.

We generate highly configurable electric fields in the \textbf{x}-\textbf{y} plane using a set of six in-vacuum electrodes\cite{Valtolina2020,Matsuda2020,Covey2017}. To orient the induced dipole moments of the molecules, we rotate \textbf{E} by a variable angle $\theta$, where $\theta=0\degree$ corresponds to $\textbf{E}\parallel\textbf{x}$ (Fig. 1A). An electric field gradient $\nabla\left|\textbf{E}\right|$ can be applied along \textbf{y}, parallel to the direction of tight confinement in the optical lattice. Since the KRb rotational energy levels are sensitive to $\left|\textbf{E}\right|$, addressing single layers requires stabilizing the molecule position relative to the local electric field distribution; this is done by interferometrically measuring the positions of the lattice layers and electrodes, and minimizing relative displacements by adjusting the phase of the lattice\cite{SI}. By adding a fixed phase offset, the layers are displaced from the electrodes along \textbf{y} by a distance $\delta y$. Using microwave pulses, the molecules can be transferred between states $\ket{0}$, $\ket{1}$, and $\ket{2}$ (Fig. 1B). Each rotational state $\ket{N}$ develops an induced dipole moment $d_{N}$ parallel to \textbf{E}, the magnitude of which depends on $\left|\textbf{E}\right|$, leading to a state-dependent energy shift of $-d_{N}\left|\textbf{E}\right|$. At $\left|\textbf{E}\right|=1$ kV/cm, where all of the subsequent measurements are performed, the sensitivity of the $\ket{0}\leftrightarrow\ket{1}$ transition to $\left|\textbf{E}\right|$ is 40 kHz/(V/cm).

\clearpage
\begin{center}
\includegraphics[scale=0.84]{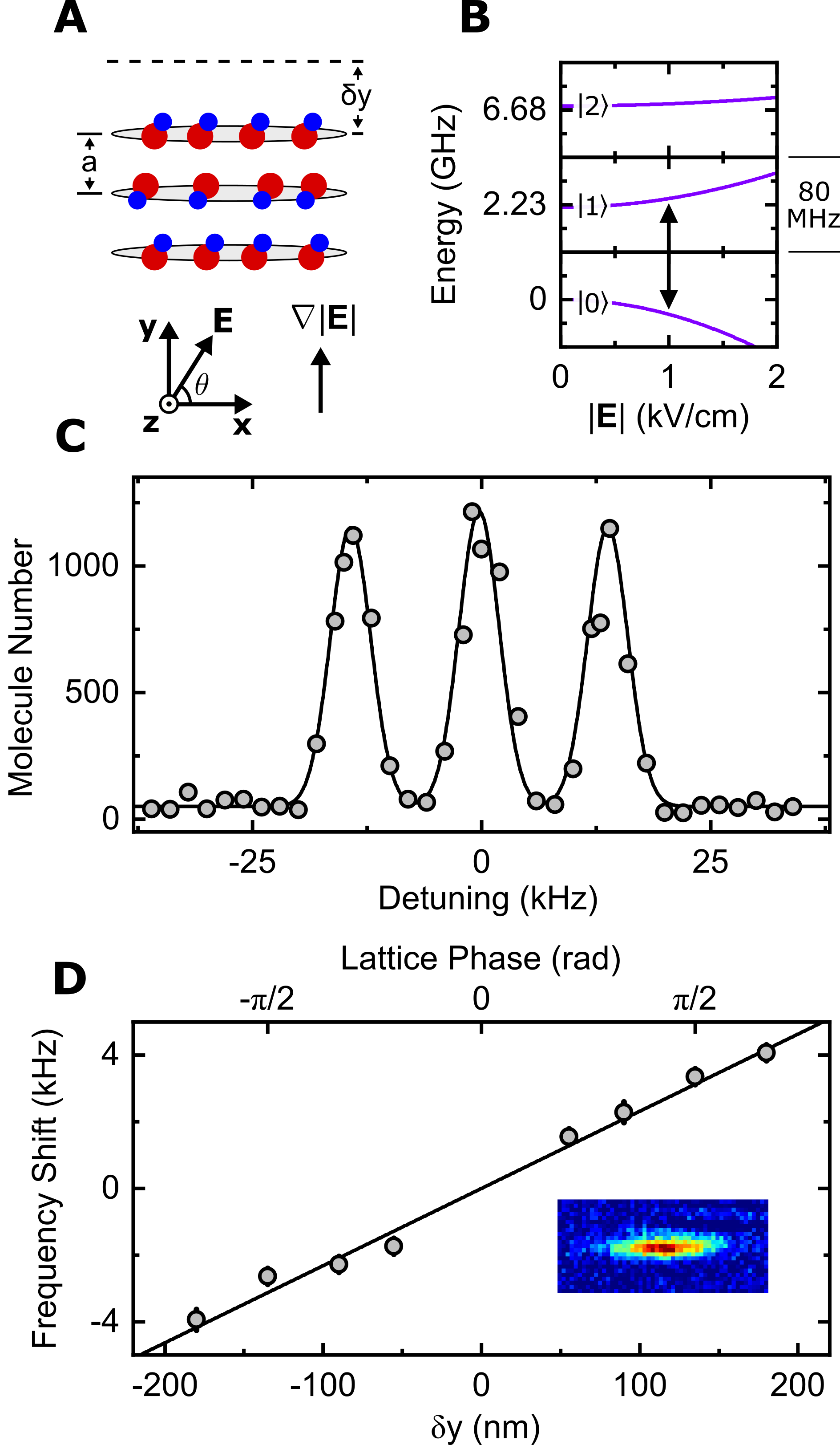}
\end{center}

\noindent {\bf Fig. 1.} Experimental configuration and individual layer addressing. \textbf{(A)} Molecules occupy two-dimensional layers in the \textbf{x}-\textbf{z} plane, separated by layer spacing $a$. The bias electric field \textbf{E} is oriented at an angle $\theta$ in the \textbf{x}-\textbf{y} plane, with an electric field gradient $\nabla \left|\textbf{E}\right|$ parallel to \textbf{y}. The lattice layers are displaced relative to the electrodes generating \textbf{E} by a distance $\delta y$. \textbf{(B)} KRb rotational structure. The arrow indicates the layer selection transition. \textbf{(C)} $\ket{0}\leftrightarrow\ket{1}$ frequency spectrum of a trilayer at $\partial_y\left|\textbf{E}\right|=6.4(2)$ kV/cm$^2$. Only three adjacent lattice layers are populated. \textbf{(D)} Center frequency shift of layer selection vs. $\delta y$. Displacements smaller than 20 nm are measured. \textbf{Inset:} Absorption image of a single layer.

\clearpage

The electric field dependence of rotational state energies enables microwave addressing of individual lattice layers. In terms of the layer spacing $a$ and the dipole moments $d_0$ and $d_1$, a field gradient $\partial_y\left|\textbf{E}\right|$ shifts the $\ket{0}\leftrightarrow\ket{1}$ transition on adjacent layers by the frequency
\begin{equation}
\Delta=\partial_y\left|\textbf{E}\right|\cdot a \cdot (d_0-d_1)/h    
\end{equation}
With a microwave pulse of sufficiently narrow spectral width, all molecules in a single layer can be addressed without a measurable effect on other layers. In addition to layer-selective addressing of the $\ket{0}\leftrightarrow\ket{1}$ transition, we have the capability to apply global microwave pulses (addressing all molecules, irrespective of $\partial_y\left|\textbf{E}\right|$) on the $\ket{0}\leftrightarrow\ket{1}$ and $\ket{1}\leftrightarrow\ket{2}$ transitions, as well as to globally remove $\ket{0}$ and $\ket{1}$ molecules with optical pulses of STIRAP light. From an initial condition of many occupied $\ket{0}$ layers, we use sequences of these microwave and optical pulses to prepare arbitrary layer configurations containing states $\ket{0}$, $\ket{1}$, and $\ket{2}$, including isolated 2D layers\cite{SI}.

To demonstrate layer selection, we prepare three adjacent layers in $\ket{1}$ and scan the frequency of an additional layer-selective $\ket{0}\leftrightarrow\ket{1}$ pulse while applying a gradient $\partial_y\left|\textbf{E}\right|=6.4(2)\text{ kV/cm}^2$. By monitoring the population transferred to $\ket{0}$ as a function of frequency, we probe the initial $\ket{1}$ distribution (Fig. 1C). We measure about 1200 molecules in each occupied layer, with adjacent layers detuned by 14 kHz. No molecules are transferred from outside the trilayer, nor at detunings halfway between occupied layers, confirming that the pulses are selectively addressing individual layers. 

By varying the layer displacement $\delta y$ and tracking the layer selection transition frequency, $\left|\textbf{E}(y)\right|$ can be extracted with high spatial resolution, far below the interlayer spacing of 540 nm or the imaging diffraction limit. To characterize this technique, we probe an applied gradient of $\partial_y\left|\textbf{E}\right|=6.4(2)\text{ kV/cm}^2$. At each of eight different $\delta y$, spanning 360 nm (corresponding to a lattice phase shift of 240$\degree$), we measure the central frequency for layer selection (Fig. 1D). Fitting the frequency shift as a function of $\delta y$, we extract $\partial_y\left|\textbf{E}\right|=\text{5.8(3) kV/cm}^2$, with a maximum offset between $\delta y$ and the line-of-best-fit of only 20 nm. These measurements demonstrate subwavelength detection of molecule distributions using electric field gradients, and high-precision electric field microscopy on nm spatial scales.

Using layer-selective addressing, we next optimize the rotational coherence in a single layer. Long-lived coherence is essential for realizing strong dipolar interactions\cite{Peter2012,Bilitewski2021}. However, inhomogeneous broadening from external electric fields and optical trapping potentials tends to limit coherence. Since molecules in 2D occupy a large spatial extent in the radial direction, electric field gradients transverse to \textbf{y} contribute a position-dependent frequency shift between rotational states (Fig. 1A). Spatial variation in the optical trap intensity contributes an additional frequency shift due to the differential ac polarizability between rotational states. The magnitude of the shift depends on $\left|\textbf{E}\right|$ as well as the angle between the quantization axis (aligned with $\textbf{E}$) and the optical lattice polarization $\boldsymbol\epsilon$\cite{Kotochigova2010}. In our apparatus, since the optical lattice polarization $\boldsymbol\epsilon$ is fixed parallel to \textbf{x}, this angle coincides with the rotation angle $\theta$ of $\textbf{E}$ (Fig. 1A). State-insensitive trapping can be achieved at the ``magic angle'' of approximately $54\degree$, where the differential ac polarizability between all rotational states with $m_N=0$ vanishes (Fig. 2A).

\clearpage
\begin{center}
\includegraphics[scale=0.84]{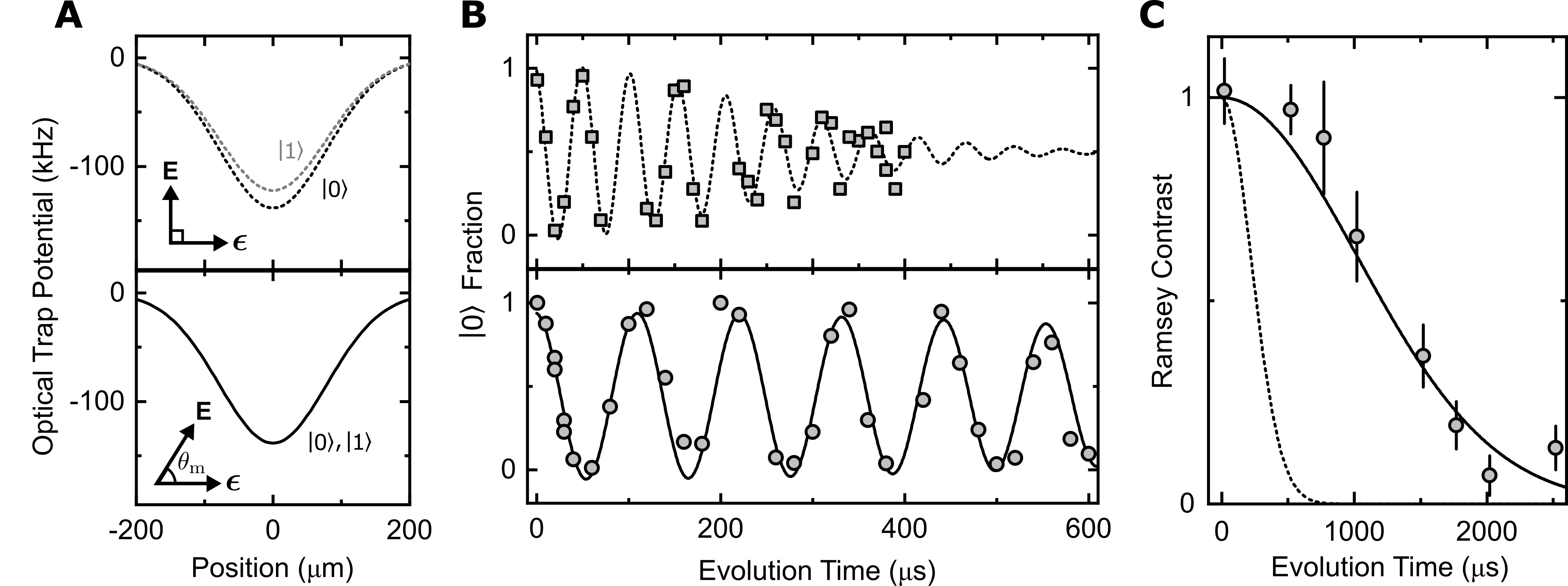}
\end{center}

\noindent {\bf Fig. 2.} Increasing rotational coherence time by rotating \textbf{E}. \textbf{(A)} Calculated optical trap potentials for states $\ket{0}$ and $\ket{1}$ in the ODT and optical lattice. The differential polarizability between rotational states depends on the angle $\theta$ between \textbf{E} and the optical lattice polarization $\boldsymbol\epsilon$, and vanishes at $\theta_\text{m}\approx 54\degree$. \textbf{(B)} Ramsey oscillations of a single layer of molecules at $\theta= 90\degree$ (top, dashed line) and $\theta = \theta_\text{m}$ (bottom, solid line). \textbf{(C)} Contrast of Ramsey fringes at long evolution times. The coherence time at $\theta=\theta_\text{m}$ (points, solid line) is increased by a factor of 5 compared to $\theta=90\degree$ (dashed line).
\clearpage

We measure the Ramsey coherence of a single 2D layer with $\partial_y\left|\textbf{E}\right|=0$ at both $\theta=90\degree$ and $\theta=\theta_\text{m}$, where we define $\theta_\text{m}$ as the electric field angle at which we measure the minimum differential polarizability \cite{SI}. For both angles, the ODT polarization is set to the magic angle with respect to \textbf{E}. Using a single layer removes possible systematics such as dipolar interactions between layers, stray electric field gradients along \textbf{y}, and layer-to-layer optical trap intensity variation. To measure the coherence decay, we prepare a single layer of molecules, use a $\pi/2$ pulse to initialize all molecules on the layer in an equal superposition of $\ket{0}$ and $\ket{1}$, hold for a variable evolution time, apply a second $\pi/2$ pulse, and simultaneously measure the population in both states. As a function of evolution time $t$, we fit the contrast envelope to the Gaussian function $e^{-t^2/\tau^2}$, where $\tau$ is the coherence time. For $\theta=90\degree$, we measure $\tau=$ 310(30) $\mu$s (Fig. 2B, top). This is consistent with simulations where the differential ac polarizability is the only mechanism causing decoherence\cite{SI}.

At $\theta=\theta_\text{m}$, little contrast decay is observed over 600 $\mu$s (Fig. 2B, bottom). At longer evolution times, the Ramsey oscillation phase is scrambled by slight changes in $\left|\textbf{E}\right|$ between experimental runs. Here, we randomize the phase between the $\pi/2$ pulses at a fixed time and compute the contrast based on the observed variance of the rotational state populations\cite{SI}. We measure $\tau=1450(80)$ $\mu$s (Fig. 2C), a factor of five improvement over $\theta=90\degree$ and exceeding the longest bulk coherence time previously observed for KRb\cite{Neyenhuis2012}. Factors that may limit the maximum achieved coherence include any remaining differential ac polarizability, residual electric field gradients, and intralayer dipolar interactions. With ms-scale coherence times and realistic experimental parameters, KRb is predicted to dynamically generate spin-squeezed states in 2D\cite{Bilitewski2021}.

The capability to prepare arbitrary layer configurations enables the realization of novel interacting systems. Here, we study a system where the rate of chemical reactions on a single 2D layer can be controlled by the presence of adjacent layers\cite{Pikovski2011}. The dynamics of molecules on multiple layers depends on a number of processes (Fig. 3A). In a single layer, molecules undergo two-body chemical reactions according to the rate equation $dN/dt=-\beta N^2$, where $N$ is the molecule number and $\beta$ is the two-body rate coefficient. KRb is fermionic, so ultracold molecules in the same internal state undergo reactions in the $p$-wave channel, with rate coefficient $\beta_p$. Molecules in different rotational states are distinguishable and therefore react in the $s$-wave channel, with rate coefficient $\beta_s$, which is typically orders of magnitude higher than $\beta_p$ due to the absence of a centrifugal barrier\cite{Idziaszek2010,Ospelkaus2010}. Molecules in separate layers in different rotational states may also exchange rotational angular momenta via long-range dipolar interactions\cite{Yan2013}, changing harmonic oscillator modes in the process ($\gamma$, Fig. 3A). Spin exchange can only occur between states of opposite parity, meaning that $\ket{0}\leftrightarrow\ket{1}$ exchange is allowed and $\ket{0}\leftrightarrow\ket{2}$ exchange is forbidden at $\left|\textbf{E}\right|=0$. Applying an electric field induces rotational state mixing, but this effect only slightly weakens the above selection rules at $\left|\textbf{E}\right|=1$ kV/cm. 

% Elastic collisions within layers can also redistribute molecules among harmonic modes, but the ratio of elastic to inelastic collisions is predicted to be smaller than one for these experimental parameters due to the small induced dipole moments\cite{Quemener2010,Valtolina2020}. As such, we neglect this effect.

Spin exchange facilitates the mixing of rotational state populations between initially spin-polarized layers, causing molecules undergoing exchange to be rapidly lost with rate coefficient $\beta_s$\cite{Pikovski2011}. In order to distinguish exchange from chemical reactions, we first measure the rate coefficients $\beta_p$ and $\beta_s$ at the temperature $T=334(30)$ nK, with $\nabla\left|\textbf{E}\right|=0$. Throughout the following, we describe bilayer and trilayer configurations according to the rotational states present in layers containing molecules: for example, ``202'' refers to a central layer containing only molecules of state $\ket{0}$ with adjacent $\ket{2}$ layers above and below, and all other layers unoccupied. To extract $\beta_p$, which in general depends on rotational state due to variation in the intermolecular potentials \cite{Kotochigova2010_2}, we prepare spin-polarized 000 (Fig. 3B, green squares), 111, and 222 trilayers and fit the number decay to the two-body loss rate equation. For $\ket{0}$ and $\ket{1}$, $\beta_p=2.99(17)\times10^{-3}$ s$^{-1}$. For $\ket{2}$, the loss rate is reduced to $\beta_p=1.78(24)\times10^{-3}$ s$^{-1}$. To extract $\beta_s$, we prepare a 111 trilayer, and apply a $\pi/2$ pulse to form an equal superposition of $\ket{0}$ and $\ket{1}$, which decoheres completely within several ms (Fig. 3B, inset). We measure $\beta_s=2.0(3)\times10^{-1}$ s$^{-1}$, nearly two orders of magnitude larger than $\beta_p$, as has been previously observed\cite{Ospelkaus2010}.

The interplay of exchange and loss is evident in layer configurations where multiple rotational states are present. In a 202 trilayer, where selection rules disallow spin exchange, the loss rate of $\ket{0}$ molecules from the central layer matches $\beta_p$ (Fig. 3B, orange diamonds). By contrast, for a 101 trilayer, the effective two-body loss rate increases by more than a factor of 10 (Fig. 3B, blue circles). The spin exchange rate depends on the density of molecules in adjacent layers, which is analogous to the dependence of the chemical reaction rate on the local molecule density. We demonstrate this effect by preparing 000, 10, and 101 layer configurations, and fitting the decay of $\ket{0}$ molecules in the central layer to the two-body loss rate equation. The fit $\beta$ scales linearly with the number of adjacent $\ket{1}$ layers (Fig. 3C). These results show the dependence of spin exchange on density and rotational state, and demonstrate tuning of the chemical reaction rate using experimental control of the layer configuration. 

\clearpage
\begin{center}
\includegraphics[scale=0.84]{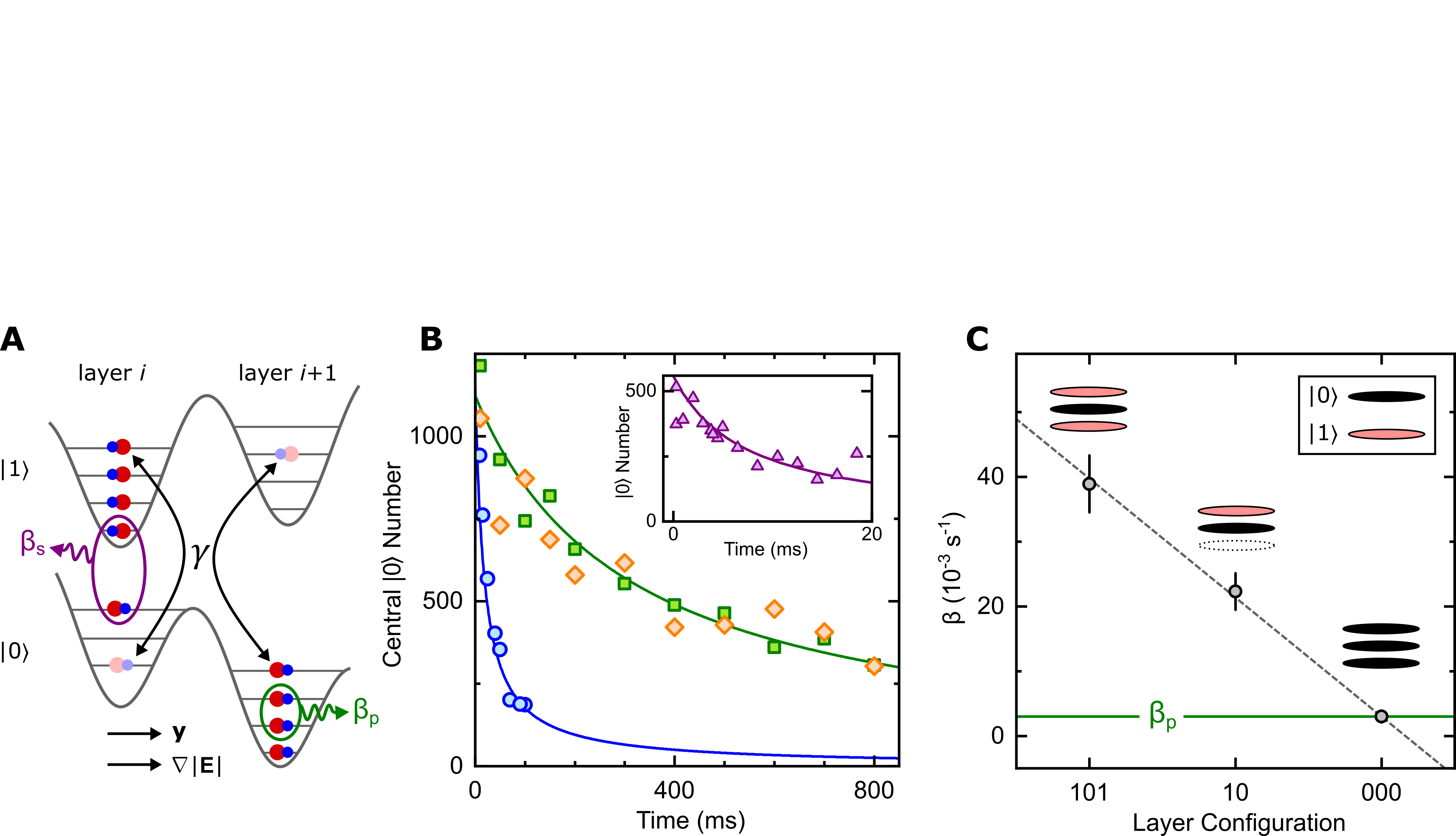}
\end{center}

\noindent {\bf Fig. 3.} Interaction and loss dynamics for molecules in 2D. \textbf{(A)} In a single layer, molecules in the same and in different rotational states undergo two-body loss with rate coefficients $\beta_p$ and $\beta_s$, respectively. Molecules in different rotational states in adjacent layers may also exchange rotational states with rate $\gamma$, potentially changing harmonic oscillator modes along the radial (\textbf{x}-\textbf{z}) direction of the trap during the exchange. \textbf{(B)} Central layer $\ket{0}$ molecule number vs. time for 000 ($\beta_p$, green squares), 101 (blue circles), and 202 (orange diamonds) trilayers. The solid lines are fits to the two-body loss rate equation. \textbf{Inset:} Loss for an equal mixture of molecules in states $\ket{0}$ and $\ket{1}$ ($\beta_s$). \textbf{(C)} Density dependence of spin exchange. $\beta$ fit to the loss of $\ket{0}$ from the central layer of a trilayer scales linearly with the number of adjacent layers containing $\ket{1}$. The solid green line indicates $\beta_p$.\\

\clearpage

To extract the spin exchange rate quantitatively, we describe $N_i^\sigma$, the molecule number in layer $i$ and in rotational state $\ket{\sigma}$, by a set of coupled differential equations including the aforementioned loss and exchange processes:
\begin{equation}
\frac{dN_i^\sigma}{dt}=-\beta_p N_i^\sigma N_i^\sigma -\beta_s N_i^\sigma N_i^{\sigma'}+\gamma \sum_{k=i\pm1}\left(N_i^{\sigma'}N_k^\sigma-N_i^\sigma N_k^{\sigma'}\right)
\end{equation}
$\sigma\ne\sigma'$ are the two rotational states participating in the dynamics. The first two terms represent intralayer two-body loss, with rates $\beta_p$ and $\beta_s$ for spin-polarized and spin-mixed molecules, respectively. The third term represents spin exchange, which depends on the molecule populations in different rotational states in adjacent layers, occurring with rate constant $\gamma$ (Fig. 3A). $\gamma$ is an effective parameter describing the spin exchange, averaged over all molecules and over the full duration of the measurement.

Since spin exchange is a resonant process, adding an energy offset between adjacent layers suppresses its rate. To probe the energy spectrum of exchange, we add a variable gradient $\partial_y\left|\textbf{E}\right|$ (Fig. 3A). The total change in electric potential energy when molecules in adjacent layers exchange rotational states is $h\Delta$ (Eq. 1), which is equivalent to the shift in microwave transition energy between adjacent layers (Fig. 1C). For states $\ket{0}$ and $\ket{1}$, $\Delta=$ 14 kHz at $\partial_y\left|\textbf{E}\right|=6.4$~kV/cm, the gradient used for layer selection.

We measure the spin exchange rate $\gamma$ as a function of $\Delta$ in 101 and 202 trilayer configurations (Fig. 4A), with $\theta=90\degree$ and at $T=334(30)$ nK. For 202, the measured $\gamma$ is consistent with zero spin exchange and does not depend on $\Delta$. For 101, however, the peak exchange rate is $\gamma=7.0(6)\times 10^{-3}\text{ s}^{-1}$, more than two times $\beta_p$. Strikingly, $\gamma$ remains non-zero for large $\Delta$, with a Lorentzian fit to $\gamma(\Delta)$ having a full width at half maximum (FWHM) of 6.4(6) kHz. This energy scale vastly exceeds the dipolar interaction energy between two molecules: at a separation of 540 nm, the rate of spin exchange between molecules in $\ket{0}$ and $\ket{1}$ is only 100~Hz\cite{Yan2013}. 

Thermal energy contributes to the broad linewidth. To compensate $\Delta$ and conserve energy during spin exchange, molecules must change harmonic modes (Fig. 3A), the initial occupation of which is determined by the temperature. Qualitatively, this mechanism gives insight into the scaling of $\gamma$ with $T$ and $\Delta$. At low temperatures, no spin exchange can occur when $\Delta$ greatly exceeds the thermal energy. For small $\Delta$, however, the spin exchange rate is enhanced because of the high average occupation and strong dipolar coupling of low-lying harmonic modes. At high temperatures, the situation is reversed: high-lying modes are occupied, allowing exchange even at large $\Delta$, but the peak exchange rate on resonance is suppressed. These effects suggest that increasing the temperature should broaden the spin exchange linewidth.

\clearpage
\begin{center}
\includegraphics[scale=0.84]{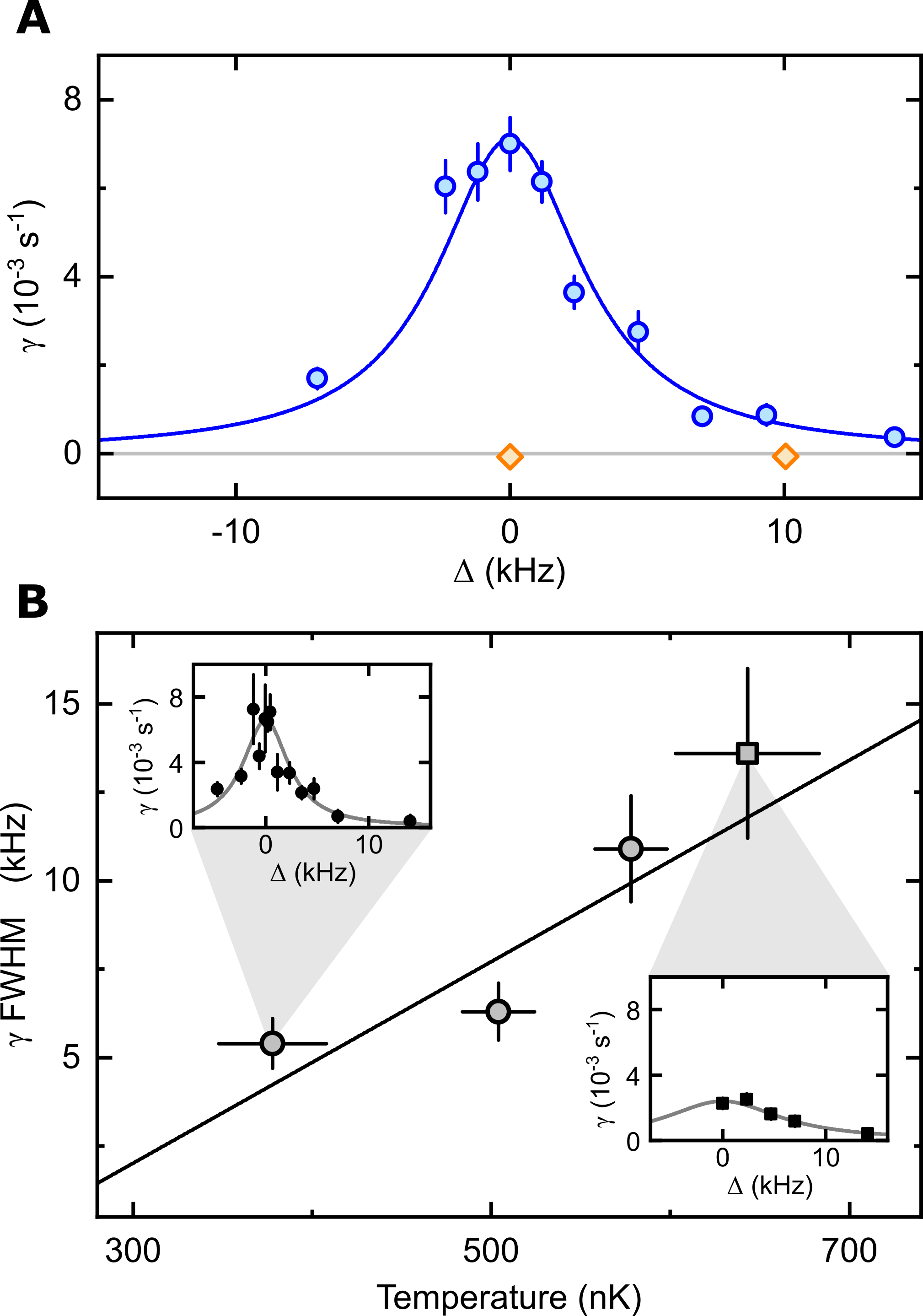}
\end{center}

\noindent {\bf Fig. 4.} Dependence of spin exchange on interlayer detuning $\Delta$, rotational state, and temperature. \textbf{(A)} Spin exchange rate vs. $\Delta$, with $\theta=90\degree$ and $T=334(30)$ nK, for 101 (blue circles) and 202 (orange diamonds) trilayers. The solid line is a Lorentzian fit to extract the FWHM. The point displayed at $\Delta=0$ kHz is the weighted average of measurements at $\Delta=0$ and $\pm0.12$ kHz. \textbf{(B)} Temperature dependence of spin exchange linewidth, with $\theta=\theta_\text{m}$. The solid line is a linear fit for the temperature range shown. \textbf{Insets}: $\gamma$ vs. $\Delta$ at $T=378(30)$ nK (circles, upper) and $T=643(40)$ nK (squares, lower).

\clearpage

To quantify the temperature dependence, we repeat the measurement of $\gamma(\Delta)$ with a 101 trilayer at four temperatures between 370-650 nK (Fig. 4B). We set $\theta=\theta_\text{m}$ in order to eliminate possible broadening due to varying trap potentials between rotational states (Fig. 2A). Here the strength of dipolar interactions between harmonic modes is slightly altered because the dipole moments are rotated relative to the plane of motion\cite{SI}. At the lowest temperature we measure a FWHM of 5.4(7) kHz, only slightly narrower than at $\theta=90\degree$. At the highest temperature, by contrast, the peak $\gamma$ is reduced and the FWHM more than doubles to 13.6(2.4) kHz. To provide physical insight into the temperature dependence, we use a simple two-particle model of molecules in adjacent layers interacting via dipolar spin exchange\cite{SI}. Weighting by the thermal mode occupation, we find qualitative agreement with the observed trend. In addition to thermal energy, effects that may contribute to broadening include many-body interactions, where multiple molecules participate jointly in the spin exchange process, and intralayer dipole-dipole interactions. Both of these mechanisms should only weakly affect the temperature scaling, due to the relatively low molecule density and small dipole moments at $\left|\textbf{E}\right|=1$ kV/cm. 

We have demonstrated experimental control over spin exchange and chemical reactions in two-dimensional systems of ultracold molecules, enabled by subwavelength addressing of individual lattice layers. These results provide a general method for layer-resolved state preparation and imaging of polar molecules, facilitating the study of many-body phases and non-equilibrium dynamics in long-range interacting systems with reduced dimensionality.\\

\noindent\textbf{Acknowledgments:} We thank Sam R. Cohen, Luigi De Marco, and Giacomo Valtolina for experimental contributions, and Yu Liu and Christian Sanner for helpful discussions.  Funding for this work is provided by ARO MURI, AFOSR MURI,
DARPA DRINQS, NSF QLCI OMA–2016244, NSF Phys-1734006, and NIST.\\

\noindent\textbf{Author contributions:} The experimental work and data analysis
were done by W.G.T., K.M., J.-R.L., C.M., A.N.C. and J.Y. The two-particle exchange model was developed by T.B. and A.M.R. All authors contributed to interpreting the results and writing the manuscript.\\

\noindent\textbf{Competing interests:} The authors declare no competing interests.\\

\bibliography{notes}

\bibliographystyle{Science}

\clearpage

\section*{Supplementary Materials}
\vspace{0.15 in}

\subsection*{Layer Selection Protocol}

Using a combination of microwave and optical pulses, we prepare arbitrary layer configurations of rotational states $\ket{0}$, $\ket{1}$, and $\ket{2}$. In a gradient $\partial_y\left|\textbf{E}\right|$, rotational transitions are detuned between layers, so pulses with a narrow spectral width (``layer-selective") can be used to address individual layers while pulses with a broad width (``global") address all layers. The sequence for preparing a 101 trilayer is as follows (Fig. S1):
\begin{enumerate}
  \itemsep0em
  \item Initial condition: many occupied layers in state $\ket{0}$
  \item Three subsequent layer-selective $\ket{0}\leftrightarrow\ket{1}$ pulses, detuned by $0, \pm14$ kHz from the resonant frequency of the central layer, to transfer three layers to $\ket{1}$
  \item Global $\ket{1}\leftrightarrow\ket{2}$ pulse, to transfer the three selected layers to $\ket{2}$
  \item Global STIRAP pulse, to remove the remaining molecules in $\ket{0}$
  \item Global $\ket{1}\leftrightarrow\ket{2}$ pulse, to return layers to $\ket{1}$
  \item Layer-selective $\ket{0}\leftrightarrow\ket{1}$ pulse, to transfer the central layer to $\ket{0}$\\
\end{enumerate}

\begin{center}
\includegraphics[scale=0.84]{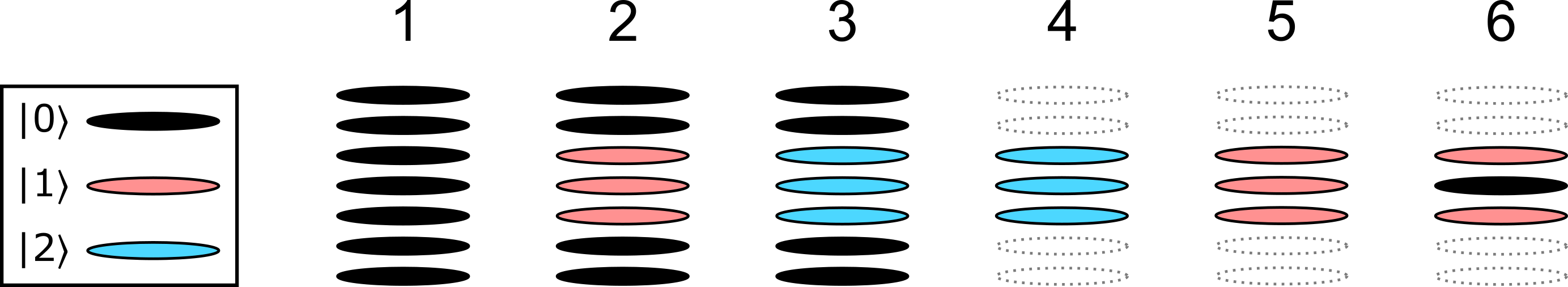}
\end{center}
\noindent {\bf Fig. S1.} Pulse sequence for layer selection.\\

We find experimentally that the STIRAP pulse removes both $\ket{0}$ and $\ket{1}$ molecules but not $\ket{2}$, necessitating the involvement of $\ket{2}$ in layer selection. Using $\ket{2}$ also  enables simultaneous imaging of states $\ket{0}$ and $\ket{1}$ in each experimental run, since $\ket{1}$ can be safely shelved in $\ket{2}$ while $\ket{0}$ is imaged using STIRAP. State-resolved imaging allows normalization of the molecule number in each state against the total molecule number, reducing sensitivity to overall number fluctuations.

\subsection*{Lattice Phase Stabilization}

Figure S2 shows a schematic of the experimental apparatus. Electric fields are generated by two transparent plates and four rods (only the plates are shown below) that are rigidly mounted together. The optical lattice is formed by two beams that enter the electrode assembly tilted at approximately $10\degree$ with respect to \textbf{y}. Beam 1 directly intersects the position of the molecules, while beam 2 first reflects off of a dichroic mirror beneath the electrodes before interfering with beam 1. Both beams are derived from a common source immediately prior to entering the experimental apparatus, and therefore initially have a static phase difference; the phase of beam 1 can be shifted by a variable amount $\Delta\phi$ by adjusting its path length.\\

\begin{center}
\includegraphics[scale=0.84]{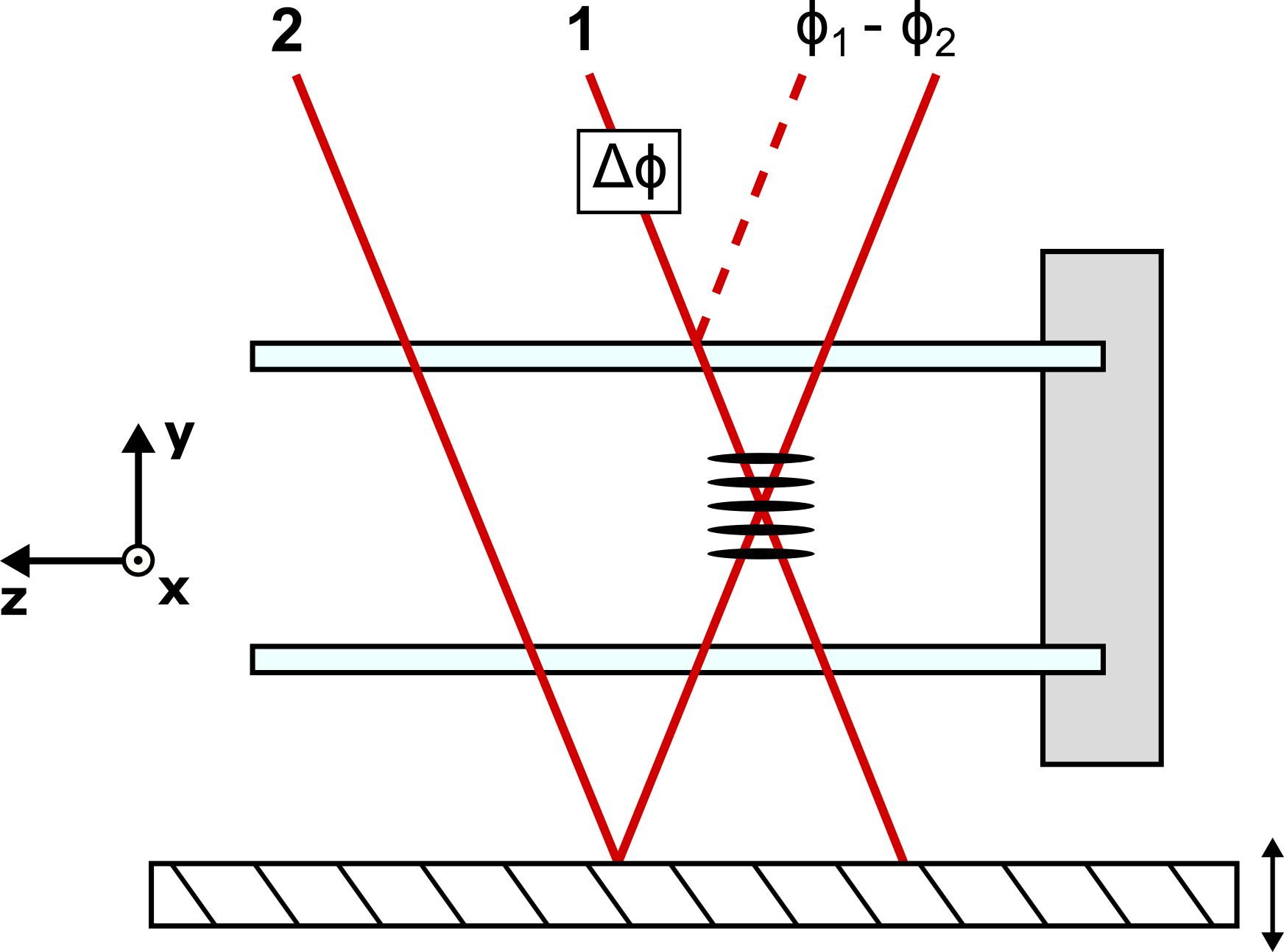}
\end{center}
\noindent {\bf Fig. S2.} Experiment schematic, showing the electrodes (light blue), electrode mount (gray), dichroic (striped), and lattice beams (red).\\

The electrode mount and dichroic may move relative to each other and relative to a space-fixed position reference. Since the electrodes generate the field gradient for layer selection, it is essential to stabilize the position of the lattice layers to the electrodes and to cancel shifts due to the motion of the dichroic. We achieve this by measuring $\phi_1-\phi_2$, the phase difference between beam 2 after the dichroic and a weak reflection of beam 1 from the upper electrode plate (Fig. S2, dashed red line). By feeding back to $\Delta \phi$ to fix $\phi_1-\phi_2$, we fix the lattice phase with respect to the electrodes. To control $\delta y$ (Fig. 1D), the displacement of the lattice from the electrodes, we vary $\phi_1-\phi_2$.

\subsection*{Ramsey Coherence Simulation}
We perform a simple simulation to estimate the decoherence rate of Ramsey oscillations due to the differential polarizability between states $\ket{0}$ and $\ket{1}$. With $\left|\textbf{E}\right|=1$ kV/cm and $\theta=90\degree$ (Fig. 2A, upper panel), the polarizabilities of $\ket{0}$ and $\ket{1}$ differ by approximately 20\% at 1064 nm. Since the optical trap intensity varies over the molecule distribution, separated molecules undergo Ramsey precession at slightly different frequencies, causing dephasing.

To simulate this effect, we initialize a 2D distribution of molecules using the temperature and radial trap frequencies corresponding to the experimental conditions. Each molecule is assigned a position-dependent frequency shift $\Delta\nu = \eta\, \Delta \alpha\, I(x,z)$, where $\Delta \alpha$ is the differential polarizability and $I(x,z)$ is the optical lattice intensity at the molecule position. $\eta$ is a scaling factor on $\Delta \alpha$, accounting for changes in the polarizabilities as a function of the relative angle $\theta$ between the electric field and the optical lattice polarization: for example, at $\theta=90\degree$, $\eta=1$, while at $\theta=\theta_\text{m}$ (where the polarizabilities of both states match), $\eta\approx0$. The total Ramsey oscillation signal, which models the experimental measurement, is obtained by summing the signal from each molecule. By fitting the decay envelope of the simulated signal to the form $e^{-t^2/\tau^2}$ (Fig. S3A), where $t$ is the evolution time and $\tau$ is the coherence time, the effective decoherence rate can be extracted as a function of $\eta$ (Fig. S3B).

For $\eta=1$, the simulation predicts a decoherence time of 400 $\mu$s, compared to the measured value of 310(30) $\mu$s (Fig. 2B, upper panel). In general, the simulation results represent an upper bound on $\tau$, since other mechanisms such as electric field gradients and molecule-molecule interactions may contribute to decoherence. Due to molecular motion during the evolution time, each molecule actually experiences a time-varying frequency shift, instead of the static shift simulated here. However, since the maximum Ramsey evolution time in experiment (2.5 ms) is small compared to the period of oscillation in the optical trap (20 ms), we neglect this effect.\\

\begin{center}
\includegraphics[scale=0.84]{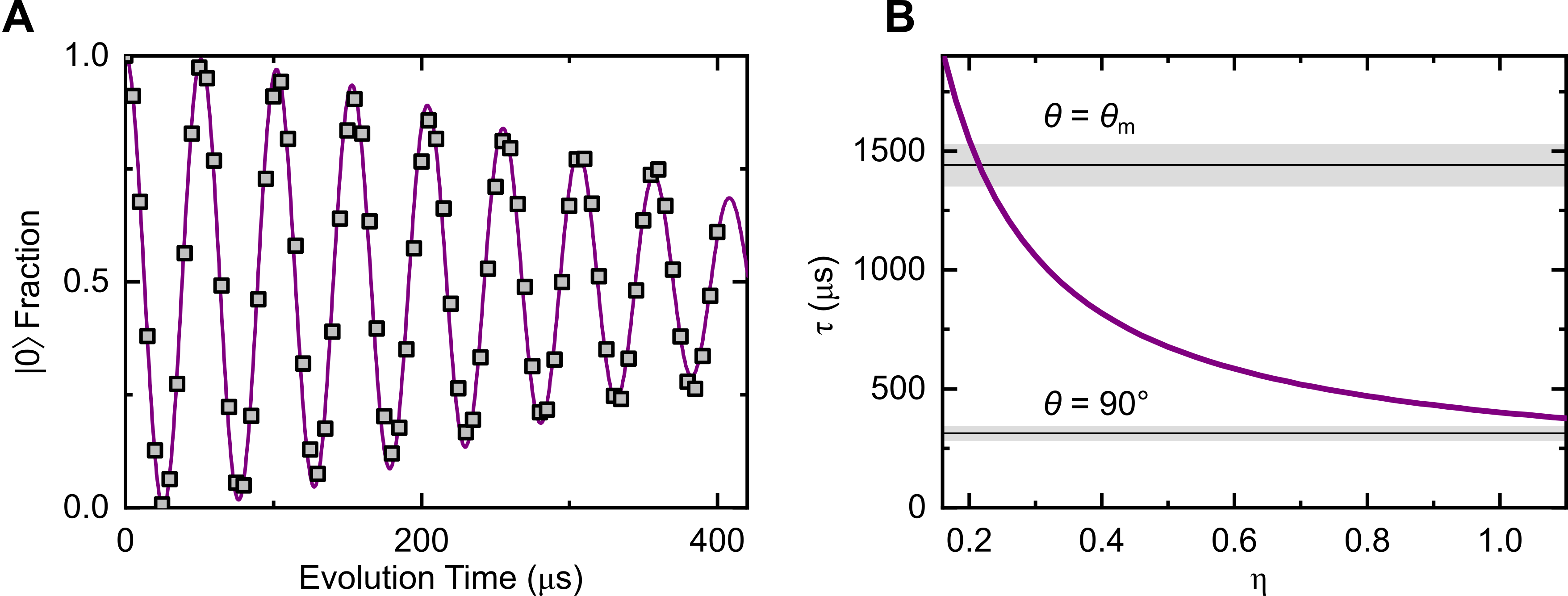}
\end{center}

\noindent {\bf Fig. S3.} Simulated decoherence. \textbf{(A)} Ramsey oscillations at $\eta=1$ ($\theta=90\degree$). The points are simulation results and the solid line is a fit to a Gaussian envelope (Eq. 1). \textbf{(B)} Coherence time vs. differential polarizability. The shaded areas represent experimental results at $\theta=90\degree$ and $\theta=\theta_\text{m}$.\\

To minimize $\Delta\alpha$ experimentally, we rotate the electric field angle $\theta$ to minimize the frequency shift of the $\ket{0}\leftrightarrow\ket{1}$ transition as a function of optical trap intensity. At the angle where this shift is minimized, which we denote $\theta_\text{m}$, we measure $\tau=1450(80)$ $\mu$s (Fig. 2C). This corresponds to $\eta\approx0.2$ in the simulation (Fig. S3B), or a factor of five reduction in $\Delta \alpha$, although $\tau$ is likely limited by gradients or interactions here.

\subsection*{Measuring Long-Time Coherence}

At short Ramsey evolution times $t$ (Fig. 2B), we fit the fraction $f$ of molecules in state $\ket{0}$ to the functional form
\begin{equation}
    f(t)= \frac{1}{2}+\frac{e^{-t^2/\tau^2}}{2}\cos(2 \pi \nu t +\phi)
\end{equation}
where $\nu$ is the precession frequency, $\phi$ is a fixed phase, and $\tau$ is the coherence time. For $t$ longer than about 600 $\mu$s, however, variation in $\left|\textbf{E}\right|$ changes $\nu$ slightly between experimental runs. 

To extract the Ramsey contrast $\mathcal C(t)$ at long times, we instead use the following method: at a fixed $t$, we randomize the relative phase of the two Ramsey $\pi/2$ pulses, and measure $f$. We repeat this process between 10 and 20 times and calculate $\sigma_{f}(t)$, the standard deviation of $f$. $\sigma_{f}(t)$ decreases at large $t$ as the molecules decohere. At $t\gg\tau$, $\sigma_{f}(t)$ is non-zero even in the absence of coherence because of imaging noise. Denoting these residual number fluctuations $\sigma_0$, the contrast is calculated by the expression
\begin{equation}
\mathcal C(t)=2\sqrt{2}\sqrt{\sigma_{f}^2(t)-\sigma_0^2}    
\end{equation}
where the prefactor is the calculated standard deviation of $f$ for $\mathcal C=1$ (that is, $\mathcal C$ at $t=0$), in the absence of imaging noise. We verify this factor by measuring $\mathcal C$ at $t=21$ $\mu$s, much shorter than the decoherence time $\tau$, and find $\mathcal C=1.02(8)$. By fitting $\mathcal C(t)$ to the form $e^{-t^2/\tau^2}$, we can extract $\tau$ (Fig. 2C).

\subsection*{Spin Exchange Model}

We estimate the rate of spin exchange between layers using a Fermi's golden rule (FGR) treatment of the two-particle problem with one molecule in each layer. The layers are separated by a lattice spacing $a=540$ nm, and within each layer the particles are harmonically confined with trapping frequencies $(\omega_x, \omega_y, \omega_z) = 2\pi \times (45,17\,000,45)$ Hz. Note that here for simplicity we have taken the potential to be radially isotropic in the \textit{x-z} plane. The two molecules interact via long-range dipolar spin-exchange 
\begin{equation}
 \hat V_{dd} ={\mathcal V} (\vec{r}) \left(\hat\sigma_1^{+} \hat\sigma_2^{-} + \hat\sigma_1^{-} \hat \sigma_2^{+} \right)
\end{equation}
where ${\mathcal V (\vec{r})} = \frac{\mu_{\downarrow\uparrow}^2}{ 4 \pi \epsilon_0 r^3} \left( 1 - 3 (\hat{\varepsilon}\cdot \hat{r})^2\right)$, $\hat{\varepsilon} = \textbf{E}/|\textbf{E}|= (\cos(\theta), \sin(\theta), 0)$ is the orientation of the electric field, $\hat{r} = (x,y,z)/\sqrt{x^2+y^2+z^2}$ is the relative coordinate between two molecules, and $\hat\sigma^{+}$, $\hat \sigma^{-}$ are Pauli raising and lowering operators acting on the two relevant rotational molecular states, $\ket{\downarrow} = \ket{N=0,m_N=0}$ and $\ket{\uparrow} = \ket{N=1,m_N=0}$. We assume that the electric field is oriented at the ``magic angle" ($\theta\approx 54^{\circ}$) where the differential polarizability between $\ket{\uparrow}$ and $\ket{\downarrow}$ vanishes, so that the two states have the same trapping frequencies. When the dipolar interactions flip the internal levels of the two molecules, $\ket{ \uparrow_1 \downarrow_2} \to \ket{\downarrow_1 \uparrow_2}$, there is an extra energy cost $h\Delta $ associated with the energy splitting between the two rotational levels in each layer generated by an applied electric field gradient.

In the basis of $\ket{\Uparrow}= \ket{\uparrow_1 \downarrow_2}$ and $\ket{\Downarrow}=\ket{\downarrow_1 \uparrow_2}$, the Hamiltonian for the relative motion reads
\begin{equation}
\begin{split}
 { \hat {\mathcal H}} =& -\frac{\hbar^2}{2M} \Big (\frac{\partial^2}{\partial x^2}+\frac{\partial^2}{\partial y^2} +\frac{\partial^2}{\partial z^2}\Big )\otimes \mathbf{1}+
\frac{M}{2} \Big(\omega^2 (x^2+z^2) + \omega_y^2 y^2\Big)\otimes \mathbf{1}\\
&+ \frac{h\Delta}{2} \hat\sigma_z + {\mathcal V}(\sqrt{2} x,\sqrt{2} y-a,\sqrt{2}z) \hat\sigma_x
\end{split}
\end{equation}
where $x=(x_1-x_2)/\sqrt{2}$, $y=(y_1-y_2+a)/\sqrt{2}$, and $z= (z_1-z_2)/\sqrt{2}$, $\omega=\omega_x=\omega_z$ is the radial trap frequency, and $M$ is the KRb mass. The quantum state of the relative motion can be specified by $\ket{n_r,m,n_y}$, where $n_r,m$ are the \textit{x-z} plane radial and angular momentum oscillator quantum numbers, and $n_y$ is the $y$ oscillator quantum number. 

The FGR predicts a spin exchange rate for a given initial state $\ket{n_r,m,n_y}$ of:
\begin{equation}
 K_{\mathrm{FGR}}(n_r,m,n_y) =\frac{ \pi}{\omega} \sum_{n_r^{\prime},m^{\prime},n_y^{\prime}} \left|V^{mm^{\prime}}_{n_r,n_r^{\prime},n_y,n_y^{\prime}}\right|^2 \delta \big( 2(n_r-n_r^{\prime}) +(n_y - n_y^{\prime}) \omega_y/\omega + (|m|-|m^{\prime}|) - 2\pi \Delta/\omega \big)\\
\end{equation}
with $V^{mm^{\prime}}_{n_r,n_r^{\prime},n_y,n_y^{\prime}} = \braket{n_r,m,n_y|\mathcal{V} (\sqrt{2}x,\sqrt{2}y-a, \sqrt{2}z)| n^{\prime}_r,m^{\prime},n^{\prime}_y} $.
We then average over all thermally occupied states to obtain the total exchange rate
\begin{equation}
 \bar{K} = \sum_{n,m,n_y} \rho(n_r,m,n_y) K_{\mathrm{FGR}}(n_r,m,n_y)
 \label{eq:FGR}
\end{equation}
where $ \rho(n,m,n_y)$ describes the thermal occupation of the relative motional states. To compare to the experimental parameter $\gamma$, extracted by measurements on a trilayer, we sum $\bar{K}$ over the layer above and below the central layer of interest.

To satisfy energy conservation, an initial radial mode $n_r$ is dominantly coupled by dipolar interactions to final modes with $n_r' = \lfloor n_r +(n_y - n_y^{\prime}) \omega_y/(2\omega) + (|m|-|m^{\prime}|)/2- \pi\Delta/\omega \rfloor$. When $\Delta\ne0$, molecules must undergo an inelastic (mode-changing) collision while exchanging rotational states. The constraint $n_r' \ge 0$ implies that energy-conserving spin exchange involving the initial state $\ket{n_r,m,n_y}$ is disallowed for sufficiently large $\Delta$. The spin exchange rate decreases with $\Delta$ since large changes in harmonic modes are suppressed by the interaction matrix elements. Furthermore, the peak spin exchange rate on resonance ($\Delta=0$) will decrease with increasing temperature, since higher radial oscillator modes are more delocalized and therefore the matrix elements involving these modes are smaller. The FGR treatment makes several simplifying assumptions, including neglecting off-resonant couplings, intralayer interactions, and particle losses, but provides a qualitative picture of the scaling of spin exchange with temperature and $\Delta$.

Figure S4 shows the FGR predictions for the FWHM of $\gamma(\Delta)$ and the peak spin exchange rate as a function of temperature, in comparison to experimental measurements. We calculate that the linewidths increase (Fig. S4A) and the peak rates decrease (Fig. S4B) with increasing temperature, in qualitative agreement with the trends observed in experiment. In the quantum degenerate regime, the linewidth is predicted to saturate near the Fermi energy $E_F$, and the overall exchange rate is predicted to increase significantly, highlighting the potential of quantum degenerate gases in confined geometries.\\

\begin{center}
\includegraphics[scale=0.84]{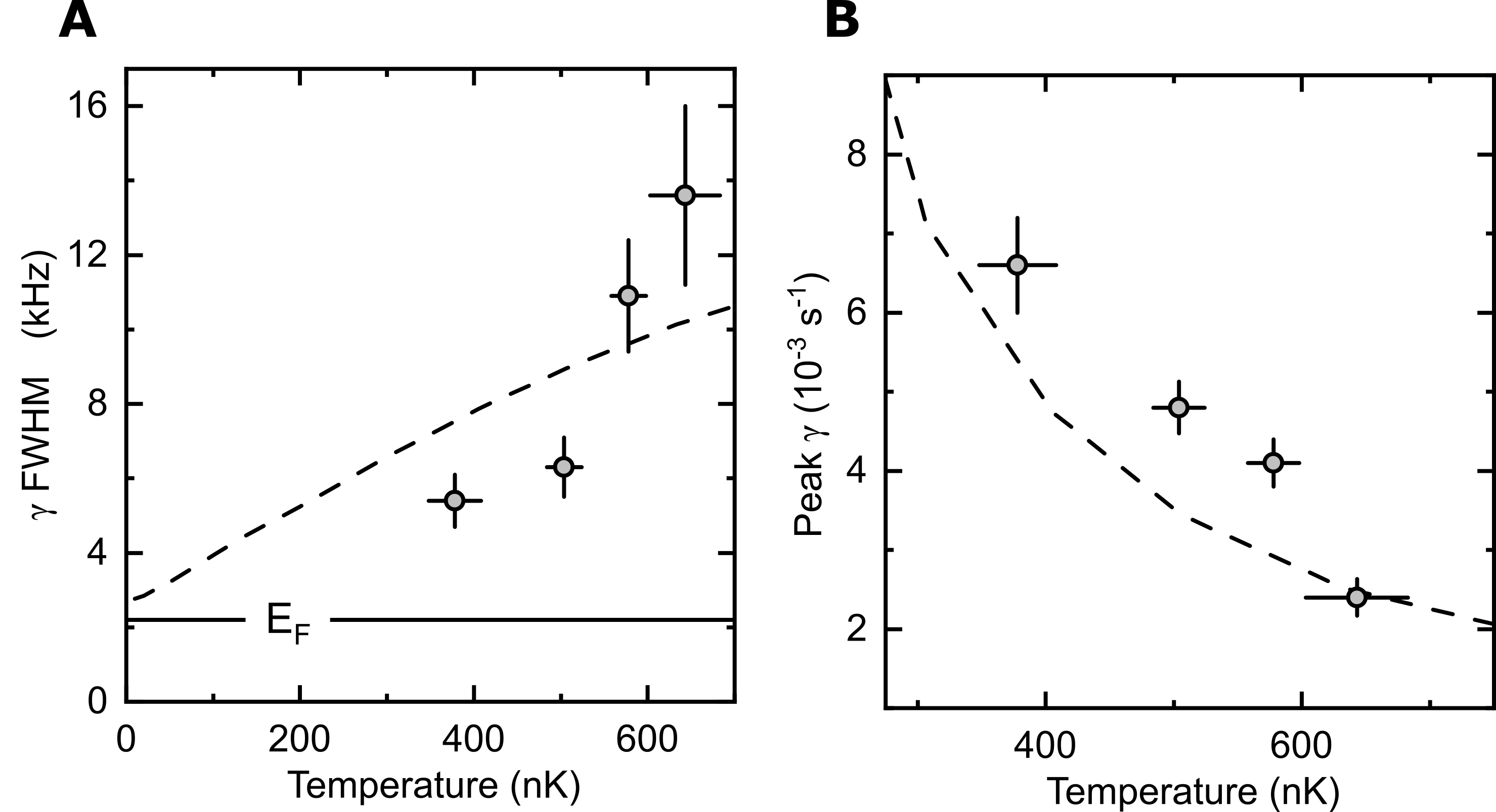}
\end{center}

\noindent {\bf Fig. S4.} Comparison of FGR theory results (Eq. 8) and experimental measurements. \textbf{(A)} FWHM of $\gamma(\Delta)$ as a function of temperature. The dashed line is the theory calculation, the points are experimental data, and the solid line represents the Fermi energy $E_F$. \textbf{(B)} Comparison of peak $\gamma$ between experiment (points) and theory (line) as a function of temperature.

\end{document}